\documentclass[conference]{IEEEtran}
\IEEEoverridecommandlockouts
\usepackage{cite}
\usepackage{amsmath,amssymb,amsfonts}
\usepackage{algorithmic}
\usepackage{graphicx}
\usepackage{textcomp}
\usepackage{xcolor}
\def\BibTeX{{\rm B\kern-.05em{\sc i\kern-.025em b}\kern-.08em
		T\kern-.1667em\lower.7ex\hbox{E}\kern-.125emX}}
\begin{document}

 \title{Health Monitoring in Smart Homes Utilizing Internet of Things}

\author{\IEEEauthorblockN{Lauren Linkous, Nasibeh Zohrabi, Sherif Abdelwahed}
\IEEEauthorblockA{Department of Electrical and Computer Engineering \\
Virginia Commonwealth University \\
Richmond, Virginia 23220 \\
Emails: linkouslc@vcu.edu, zohrabin@vcu.edu, sabdelwahed@vcu.edu}}
\maketitle
	\begin{abstract}
		In recent years the concept of the Internet of Things (IoT) has evolved to connect commercial gadgets together with the medical field to facilitate an unprecedented range of accessibility. The development of medical devices connected to internet of things has been praised for the potential of alleviating the strain on the modern healthcare system by giving users the opportunity to reside in the home during treatment or recovery. With the IoT becoming more prevalent and available at a commercial level, there exists room for integration into emerging, intelligent environments such as smart homes. When used in tandem with conventional healthcare, the IoT offers a vast range of custom-tailored treatment options. This paper studies recent state-of-the-art research on the field of IoT for health monitoring and smart homes, examines several potential use-cases of blending the technology, and proposes integration with an existing smart home testbed for further study. Challenges of adoption and future research on the topic are also discussed.
	\end{abstract}
	
	
	\section{Introduction}
	Healthcare is an enormous part of modern life, and as technology develops, so do the opportunities for care. The strain on the current healthcare system caused by an aging population and correlated chronic conditions has been well documented and explored, with consensus drawn that rigorous demand on the system has negative effects on both patients and providers\cite{morbidity, burnout}. Integrating the medical Internet of Things (IoT) with traditional practices has the potential to alleviate some of the stress placed on healthcare providers by shifting non-critical monitoring and data collection to a system capable of logging and analyzing patient data for anomalies that may indicate professional intervention is needed. 
	\par Research around the IoT in healthcare has been considerably focused on monitoring patients with specific, chronic conditions that have strong impact on the quality of life as patients age, such as those with Parkinson's disease and diabetes \cite{parkinsons, diabetes}. These particular ailments become a permanent part of a person's life and require constant attention. For example, Parkinson's disease does not have a cure, but it is possible to manage symptoms and monitor patient condition as the disease progresses. In the case of patients with diabetes, using implanted devices for continuous glucose monitoring has become commonplace and is highly recommended because these devices can provide real-time information on blood glucose concentration\cite{diabetes}. Implanted glucose monitors can also collect data over multiple days to observe trends that may only be apparent over time. However, there exist a number of non-invasive sensors that are lightweight, wireless, and can be worn by patients to help monitor for abnormal cardiac rhythms, low blood oxygen, irregular movement that may indicate fall injuries, and other symptoms. Many of these devices, and other consumer electronics, are designed to wirelessly pair with user technology such as tablets and phones to record data that can then be analyzed on a cloud server. This paper considers how medical sensing devices have been previously deployed to propose new way to use them in a smart home environment for user care.
	\par To capitalize on the versatility of wireless medical sensors, there is an increasing number of studies that look to create networks with application to a broader variety of conditions\cite{parkinsons,diabetes, af, aip}.
	These studies employ sensors in a wireless body area network (WBAN) to monitor patient health. In some cases where the devices can communicate via Bluetooth, data is forwarded to a user's smartphone \cite{baker} for processing and storage. The system considered in this paper examines a WBAN in the context of a smart home where the smartphone processing is substituted for a server capable of storing, processing, and acting on the environment based on user data.
	 \par  To further explore the idea of using medical IoT devices in the context of a smart home for the treatment and recovery of patients, this paper looks at simulating several commonly studied cases in the context of an existing smart building testbed previously presented in \cite{testbed}. The testbed is a multi-floor, multi-room physical representation of a smart building that can be tailored to  different contexts without drastically altering the underlying physical structure. The smart home system used in this paper is designed with a distributed control system that can be capitalized on by expanding the building system to include inputs from resident biometric data in order to customize the response of the home environment. With a distributed system, modification can be made to controllers on individual floors or in rooms without necessitating modification of the entire system. This allows for consideration of multiple occupants who may or may not be using wearable devices of their own. For patients with limited mobility, this can involve increasing resident comfort by adapting settings such as light intensity or temperature without affecting rooms primarily used by other residents. Additional user safety and security can be addressed by having the ability to automatically contact a caregiver or emergency services in the event of abnormal incidents where there is no user response. Safety can even be increased by introducing machine learning into the control system to learn the basic routine of a patient. For example, patients with Alzheimer's disease can have a tendency to wander off and become disoriented. In this case, the designed control system can reduce the likely-hood of patient injury or death from exposure by alerting a caregiver of unusual activity such as leaving the home during the night outside of normal routine.
	\par The structure of this paper is as follows.  Section \ref{healthcareIoT} considers the relation between the current research of medical IoT and smart buildings. Integration of WBANs to the smart home testbed for the development of test cases and further refining of current objectives are presented in section \ref{testing}. Conclusions and future research directions are presented in section \ref{conclusion}.		\begin{figure}[t]
		\centering
		\includegraphics[width=1\linewidth]{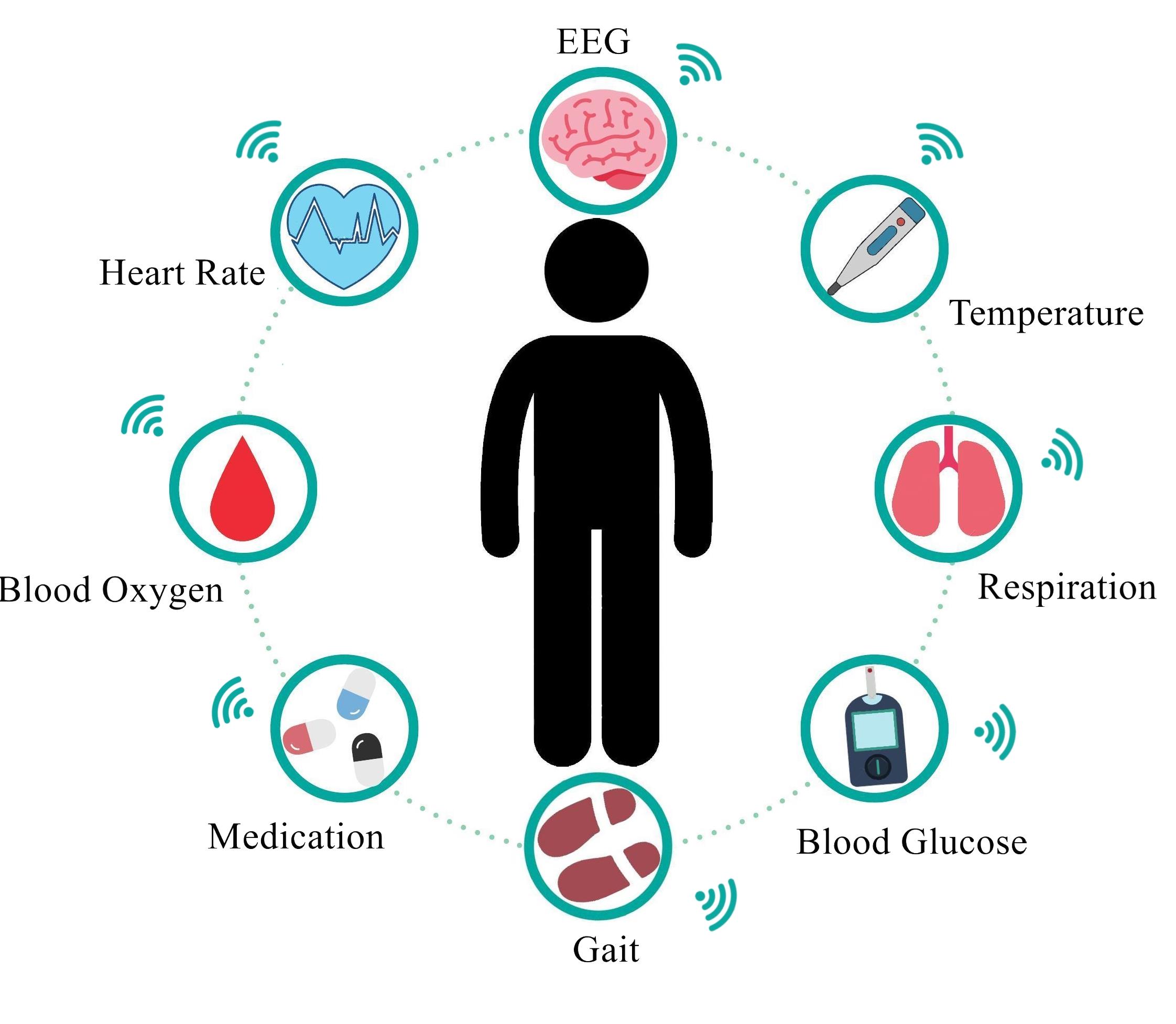}
		\caption{A medical IoT network consists of devices that can collect, analyze, and act on patient data such as brain activity (EEG), temperature, respiratory rate, blood glucose, gait, medication intake, blood oxygen, heart rate, and many more useful metrics}\vskip -3mm
		\label{fig:medicaliot}
	\end{figure}
	\section{Medical Internet of Things and Smart Homes} \label{healthcareIoT}
	\par The Internet of Things has grown to encompass any connected object embedded with a device that enables it to send and receive data.  These devices can be connected directly to the internet through an access point, or may connect to an internet enabled device through Bluetooth. The medical internet of things can consist of many devices and is customized to a user's needs. Common wearable sensors measure heart rate, blood oxygen, blood glucose, body temperature, and gait (Fig. 1). These devices can be connected either directly to the internet or to an internet enabled device such as a smartphone, laptop, or tablet.
	\par Wearable body area networks (WBANs) are a crucial component of the medical IoT \cite{baker}. The influx of commercially available devices designed to be non-invasive and generally unobtrusive to a user's daily routine has created an opportunity for tailored medical care without exorbitant equipment costs. WBANs can be single-purpose sensors such as pulse oximetry sensors that measure the level of oxygen in a user's blood, or more complex devices comparable to Fitbits, which are capable of tracking location, logging activity, and measuring heart rate. For patients in stable, non-critical conditions, monitoring using WBANs allows for recovery at-home or in the home of a caretaker instead of remaining in the hospital under observation after the completion of  primary treatment. When possible, recovery at home is often preferred due to the familiar surroundings and control that a patient has over their routine. In some countries, insurance coverage may also be a concern as patients that have been admitted for observation (who are considered outpatients) are not covered in the same way as they may be with admittance as an inpatient under a doctor's order. Moving patients out of medical facilities has the added benefit of increasing the number of free beds for patients that need more involved care and monitoring. Devices such as pill dispensers for reminding users of medication routines, personal emergency response systems (PERS), and unlit gas sensors can also be integrated into a smart home to work in tandem with wearable devices to create a more robust network for user care.
	\par Patients with chronic illnesses or that otherwise need long-term care can benefit from using WBANs to document vital information to create a more complete and detailed record of their health. This record can be useful to a medical professional by providing information or details that may not be apparent in a typical consultation visit. Documentation of abnormal conditions that show up under specific circumstances such as after intense exercise or after taking certain medications can help diagnose severe conditions before they disturb a person's daily life or result in serious consequences to a patient's health. 
	\par General monitoring of a person's well-being is possible with the medical IoT. This also applies to situations where it is not necessary to be collecting data explicitly with the intent to present it to a medical provider. Elderly persons who either do not have the means or the desire to live with a caregiver, or in an assisted facility, can benefit by having devices that can monitor falls. Moreover, ensuring the safety of users can include logging the location of elderly persons who may be in the early stages of dementia and notifying a preferred contact if the person wanders out of a regular routine or area for an extended period of time. 
		\subsection{Machine learning for the Medical Internet of Things}
	\par Medical technology has little room for error. The thin margin of error applies more strongly in cases where patients are being monitored outside of a typical medical environment. WBANs can provide an abundance of information on a person's current physical status, but they run the risk of faulty measurements, hardware failure, software issues, and any number of issues unique to the situation where they have been deployed. Incorporating machine learning to detect abnormalities in the readings of medical sensor networks has been proposed and implemented with some success \cite{anomolydetection}.  As discussed in \cite{parkinsons}, smart algorithms are effectively utilized to manage device resources and filtering large amounts of multi-dimensional data, much of which does not have correlating patterns indicative of the disease visible to human inspection. 
	\par In cases where a patient's usual readings deviate over a narrow range, large changes or spikes can indicate a health issue. Predictive modeling can be used for forecasting deviation in biometric readings, such as blood pressure, cholesterol, blood sugar, ect., from both a patient's normal baseline readings and accepted `safe' ranges. Data mining can be used with patient characteristics such as age, weight, previous medical conditions, and medication history to establish a personal baseline \cite{datamining}. When given multi-dimensional data, machine learning can be applied with more significant  effect to predict future conditions based on current trends.  Artificial neural networks (ANN) have been used to process medical data and assist in the diagnosis of some conditions \cite{machinelearning, dataminingdiabetes}. In~\cite{pasandi2019challenges}, a machine learning framework is designed to leverage IoT MAC layer technologies to deal with the heterogeneity of different IoT devices in smart home scenarios. With the application of machine learning to medical IoT it is feasible that some anomalies can be detected early, before they deviate from acceptable norms, by monitoring drift from an established patient baseline. Likewise, patient condition during recovery or changes in medication routine may be monitored for unacceptable deviations that may include low blood pressure, increased blood glucose, or a low inspired oxygen fraction.

	\subsection{Notable Features for Healthcare in Smart Homes}
	\par Smart buildings, including smart homes, are intelligent structures that utilize the Internet of Things and embedded, wired devices to increase resident comfort and reduce operation costs by efficiently managing resources. These buildings employ on-location controllers capable of collecting and analyzing data, and then controlling actuators to adjust the environment as needed to fit programmed or user-set parameters. 
	\par A medical IoT with WBANs deployed in a smart home shares several main aspects with smart buildings. However, the highest priority is given to a user's safety, followed by comfort. Notable features of a medical IoT integrated into a smart home are as follows: 
	
	\subsubsection{\textbf{User Safety}}
	\par Patient safety is the primary concern of a medical IoT network. As this network involves a person who may face potentially life-threatening health concerns, a method of contacting caregivers or emergency services is required. In cases where there is an atypical event or reading, the system will first attempt to get a response from the patient to reduce false alarms. If the patient does not respond, a caretaker should be notified. Depending on the severity of the event, emergency services may also be notified. It is possible for sensors to slip out of place and give abnormal, but not alarming, readings, so alerting a caretaker to check on sensor placement and patient health before notifying emergency services can reduce erroneous calls to emergency services. In the event of a serious issue such as a fall or cardiac arrest, or if the caretaker does not respond in a reasonable amount of time, emergency services should be contacted as soon as possible.
		\par User safety is not solely reliant on WBANs in a smart home environment. In cases where a user may be living with dementia, the risk of potential injury can be reduced by integrating commonly used appliances into the normal IoT network. If an oven or stove top has been left on for an unusual period of time, the smart home should be able to alert a resident or caregiver. With a context-aware controller, detection of such an incident can be done with the same passive infrared (PIR) sensor used to monitor movement for a security system. Furthermore, integration of medical technology in to a smart home has several advantages over short-term care facilities for personalized care. Consideration for patient routine can be worked into the network by integrating monitoring into fixed utilities and furniture. In \cite{bath} the authors integrate sensors in to a bathtub to monitor if a person becomes submerged or otherwise incapacitated while bathing. A chair with built in electrocardiogram (EKG) sensors is described in \cite{chair}, which can monitor a patient with limited mobility who may spend long periods of time in the same location. Also, if patients with dementia are prone to wandering outside of the building, a caretaker can be alerted if the user has left at unusual hours or has been outside the building for an extended period of time.

	\subsubsection{\textbf{Comfort}}
		\par Comfort in a smart building depends on the condition of the environment around the resident. Sensor placement and devices used in a WBANs have a high impact on comfort in a medical IoT network. A consideration toward patient comfort is the effort needed to engage with the devices in the network. Constant and direct engagement with devices creates inconvenience for both users and caregivers. Ideally, maintenance of devices in a WBAN should not complicate a user's routine so that the daily routine needs to be designed around constant engagement of devices in the network. Device abandonment becomes a serious issue when patients become frustrated with technology and believe that the hassle of operation is greater than the ultimate benefit \cite{deviceadoption,deviceabandonment}. If a medical device is large or otherwise designed to be non-mobile, such as a continuous positive airway pressure (CPAP) machine, the interface for operation should be straightforward. Patients who have been released to recover at home may not want or need anything beyond passive monitoring of their recovery over a short period of time. Wired devices or devices that need constant charging risk becoming a hindrance. Elderly patients living with dementia may not be capable of maintaining, cleaning, or operating complex devices.  Therefore, WBANs should be unobtrusive, with the predominant goal of not interfering with a patient's daily routine. 
	
		\subsubsection{\textbf{Patient Heath}}
		\par Medical IoT networks are not designed as a replacement for professional medical advice or care. These networks should work in tandem with a professional in order to facilitate personalized care with greater nuance than traditional consultations alone. Medical IoT in a home environment is customizable to the point where the user and caregiver are comfortable interacting with the system - if the case is that neither user nor caretaker are comfortable with or able to maintain medical IoT devices, then nothing is lost by utilizing traditional medical facilities. For users and caretakers that are comfortable with these devices, then there is the potential for freedom and autonomy that has not been previously available.
		\par In cases where insurance may not cover extended hospital stays for monitoring, WBAN devices will be available to collect data that can be reviewed by a professional. Machine learning can be implemented as a forecasting tool for potential diagnostics \cite{machinelearning}. Parkinson's disease \cite {parkinsons}, diabetes \cite{diabetes}, and Friedreich's ataxia \cite{af} have been successfully identified and monitored using wearable sensors.

	\subsubsection{\textbf{Design Flexibility}}
	    \par The adaptable nature of smart buildings offers ample opportunity for customization. In cases where the building uses a distributed control system, a system where multiple autonomous control units work towards the same goal under the supervision of a coordinating controller, individual rooms or floors can be taken offline for modification without obstructing the control units in the rest of the building. This also allows the system to be more effectively customized to individual rooms while still being able to coordinate between subsystems. 
	    For example, a controller in a kitchen can control lights, heating and air, and unlit gas monitors, while a controller located in a garage can control lights and a security system. While these systems do not need to operate independently of each other, each floor or room does not need to operate in the same ways because they have different functions. In a smart home it is conceivable that only the living area of the patient needs to react to cues from a medical IoT user's data. For smart homes where medical IoT users live with a caretaker, if the user cannot access parts of a building due to mobility issues, such as elderly patients unable to navigate stairs, then generally inaccessible areas can be separately customized for caretaker's routine, or not customized at all. This can be extended to medical care facilities, including care homes or hospices, that need to balance many patients within a single building.
	\subsubsection{\textbf{Information and User Engagement}}
		\par With many devices, user interfaces in the form of apps exist to let users view their data. Due to the number of apps and gateways needed for traditional IoT networks, many attempts \cite{gateway} have been proposed to manage device connections. The resulting dashboards, or user interface, in many ways resembles those deployed commercially for smart buildings to view environmental conditions. Allowing users to view their own data encourages positive engagement, which leads to a decrease in technology abandonment \cite{deviceadoption, deviceabandonment}. Patients can stay informed on the status of their own care by having the ability to view the data at any time.

	\section{Incorporating a Network of Medical IoT Devices into a Smart Building Testbed }
	\label{testing}

	\par  Telehealth is not a new research topic. The idea of remote patient care has been around for decades and has often been touted as a way to get consultations from specialists worldwide. However, with the recent interest in the benefits of smart buildings and the larger ecosystem of smart cities increasing, there is an interest in utilizing the connectivity in the user's home to increase the quality of life. Medical-based IoT can leverage that connectivity by extending it to personal sensor devices that allow a smart home environment to accommodate resident safety and health while still seeking to maximize comfort.  
	\par We previously designed and implemented a testbed for a smart building prototype utilizing IoT solutions to collect, analyze, and manage data from building systems \cite{testbed}. The current testbed is a multi-floor, multi-room physical representation of a smart building capable of being customized for various testing purposes. It has fully functional doors, windows, heating elements, fans, LED lights, PIR motion sensors, and cameras, placed as shown in Fig. 2. 
	The primary objective of a smart building, a cyber-physical system that adjusts to the resident's needs in a way that provides maximum comfort while minimizing operational cost, is extended here to include resident's safety and health.  For the purposes of this paper, the interaction of the building within the broader scope of a smart city is not addressed. In \cite{testbed}, the project is implemented as three units: control and data analytics, communication and data, and visualization. The building is outfitted with a distributed controller using a Raspberry Pi responsible for collecting data and controlling the actuators embedded on each floor. Communication is  realized via WiFi through HUZZAH ESP8266 modules able to communicate directly to the Raspberry Pi, which is configured as an access point to a server. Messages are transmitted via message queuing telemetry transport (MQTT) protocol. Data from the sensors can then be displayed through a dashboard deployed as a web server as part of the visualization unit. Additionally, a graphical user interface (GUI) will allow users to control aspects of the home such as temperature, lighting, and other system actuators.
	\begin{figure}[h]
		\centering
		\includegraphics[width=0.95                 \linewidth]{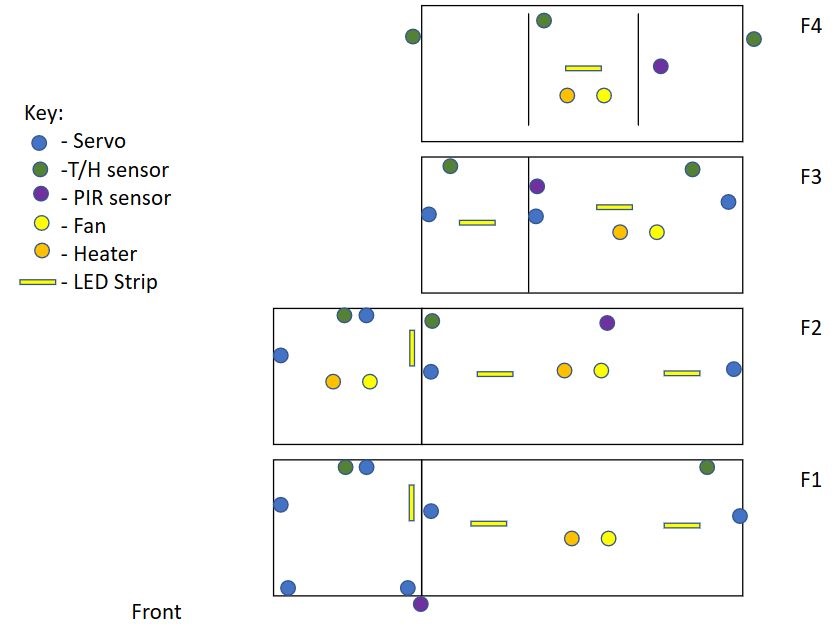}
		\caption{Layout of sensors and actuators on the four floors of the smart building testbed in \cite{testbed} }
		\label{fig:sh1}
	\end{figure}
	\par In the healthcare application, data can be collected by real people wearing medical sensors and then fed into the control system either through the server to the Raspberry Pi as a recorded dataset, or in real-time via the WiFi module if in range. For sensors that operate on Bluetooth and utilize an internet enabled device as an access point, a device can be added to the system to connect to the WiFi module or the sensor can be integrated with a secondary module capable of communicating with the WiFi module directly.
	\par Buildings are not usually designed around the idea of accommodating future health concerns, even when they are designed for accessibility. The idea of designing a structure with future capacity for further integration of smart-technology is still new.  Designing modular systems such as the sensor layout and control scheme of the testbed in \cite{testbed, eini2019distributed} permits modification beyond the original intent. This extends to medical technology implemented in the smart home environment. By utilizing a physical testbed during the design and initial implementation of a medical IoT network, unexpected behavior can be caught, isolated, and replicated to create a more robust design.  The modularity of the testbed also allows for a wider variety of flexible design in test case implementation. 
	\par For the sake of testing on controlled data, real data such as pulse, blood oxygen, respiratory rate, body temperature, and gait can be collected and stored. To simulate a health issue, anomalies can be introduced into the data to test the response of the controller. Having datasets available to replay when the controller has been modified will allow for testing for changes in response. Model predictive control~\cite{MPC, eini2019distributed} can be used to monitor biometric data, or machine learning systems such as an ANN \cite{machinelearning}  can be trained to identify trends, especially when multi-dimensional inputs are used. There may even be potential to merge an ANN with a controller depending on intended actuator impact on the environment. 
	\par Without extensive modification to the current state of the testbed as proposed in \cite{testbed}, there are several simulated use cases that can be explored:
	
		\subsubsection{\textbf{Cardiac Events}}
		\par Data to detect cardiac events can be collected from a single sensor designed to monitor heart rate, or potentially detected from a combination of heart rate monitor, blood oximeter, and blood pressure. The controller should react to sudden, abnormal spikes of activity, or complete lack of heart rate, by first attempting to get a response from the user. This can be accomplished by using audio cues to prompt a user to press a button worn on a lanyard, as is done by existing commercial products. If the patient does not respond, the controller will trigger a call to emergency services. In cases where conditions such as arrhythmia, murmurs, or palpitations exist and are documented, the controller can alert a patient or caregiver to irregular activity, but not alert emergency services unless further prompted. 
		\subsubsection{\textbf{Sudden falls}}
		\par As examined in \cite{parkinsons}, valuable data can be gained from measuring the gait of a patient. In addition to monitoring disease prognosis over time, the same sensors can be used to detect sudden movements that may be indicative of a fall. If the patient is unable to respond, or prompts for assistance, then the controller should contact emergency services.
	\subsubsection{\textbf{Conditions with limited mobility}}
		\par Limited mobility may be a result of a long-term illness, chronic condition, or accident. Awareness and capacity for the independence of a patient is conditional on the situation and may leave a patient under full-time care. These kinds of circumstances may not require emergency intervention for the duration of care. Instead, WBANs can be used in conjunction with a smart home to increase the comfort of a patient. In addition to monitoring for cardiac events, a low fraction of inspired oxygen, pulse rate, and other metrics, it is possible to measure the body temperature of a patient and adjust the environmental temperature as needed. Status of a security system may be available to a caretaker outside the home or used by a patient to monitor the movements of other residents, such as small children, living in the home.

    \par Data for conditions and situations such as those described previously can be either collected in real time (blood oxygen, heart rate, gait, testbed door open or closed) or spoofed in order to train and test the response of the smart building controller. Devices for confirming patient status after a fall or cardiac event, such as a push-button lanyard, can be implemented in the testbed. Automatically contacting a caregiver or emergency services in cases of emergency can be simulated by using researcher phones and other devices like tablets or computers. Medical data can also be displayed in a dashboard such as the one developed for our testbed in \cite{testbed}.
    
	\par Other scenarios, where WBANs and medical IoT are implemented, will require further modification to the testbed, or the introduction of (potentially simulated) 3rd party devices. Plans for aging in place will especially benefit from medical IoT in a smart home environment as the device network can be modified to accommodate different conditions that occur with age, some of which may be temporary, or have a sudden onset. Home modifications such as the bathtub in \cite{bath} and EKG measuring chair \cite{chair} are valuable with aging users who may be prone to falls or muscle weakness, and users undergoing serious treatment such a chemotherapy. It may be possible to mimic the behavior of the circuits presented in \cite{bath} and \cite{chair} on a smaller scale for triggering a response from the smart home system. Managing medication and treatment will be possible with devices such as pill dispensers connected to the network to remind patients when it is time to take pills. Mental decline and health can be monitored using ambient systems able to monitor patient routine for unusual behavior such as elderly patients not eating or not bathing\cite {mining2}. This can also be extended to apply to people with diagnosed mental conditions such as bipolarism that are non-adherent to medication routines \cite{bipolarism}, or people living with Post Traumatic Stress Disorder (PTSD), where sudden routine changes can be an indicator of suicide risk \cite{suicide, PTSD}. 
	
	\section{Conclusion and Future Work} \label{conclusion}

	\par Good health is one thing that many people take for granted. Injury, sickness, and other ailments do not discriminate and can affect any person at any time. As technology and interest in smart homes develops, so does the opportunity to integrate healthcare. The Internet of Things offers a flexible, customizable solution to not only alleviate the current strain on healthcare providers by moving routine, non-critical care into the comfort of a patient's home, but it allows patients the freedom to recover in the comfort and familiarity of their own home.
	\par There are still many challenges in the fields of IoT, smart buildings, and the evolution of telehealth. In addition to the usual issues of patient technology adoption, medical IoT runs the risk of having sensors not only be unused, but used incorrectly. Machine learning has the potential to partially mitigate this risk but relies heavily on having enough data to develop a network capable of analyzing a wide variety of sensor systems. The prevention of false alarms and other errors that can occur in intelligent systems like failing to recognize context are also issues that need to be addressed.  However, despite the challenges that exist, the field of medical IoT is continuously evolving from possibility to application.
	
	\par With the recent interest in the benefits of smart buildings and the larger ecosystem of smart cities increasing, there is an interest in utilizing the connectivity in the user's home to increase the quality of life. Medical IoT can leverage that connectivity by extending it to personal sensor devices that allow a smart home environment to accommodate resident safety and health while still seeking to maximize comfort. The testbed that we constructed previously promotes many possibilities for the development of medical device networks in a smart home setting. The modular design and distributed control system allows for a wide variety of modifications without changing the underlying physical structure of the testbed. This allows for a large number of test cases to be implemented in a physical environment for further testing and refinement of the smart home control system. Development of user-customized medical IoT networks that can be integrated beside a current control scheme will open up opportunities for modification of existing buildings without needing a complete redesign of existing systems. Furthermore, traditionally non-wearable devices, like blood pressure monitors and electrocardiogram machines, and other stationary devices such as pill dispensers can be integrated to our testbed in the future either as a third-party device or reconstructed in order to simulate system interaction with these devices.
	
	\bibliographystyle{IEEEtran}
    \bibliography{refs}

\end{document}